\documentclass{IEEEcsmag}
\usepackage[T1]{fontenc}
\usepackage[utf8]{inputenc}
\usepackage{hyperref}
\usepackage{color}
\usepackage{upmath}
\usepackage[binary-units=true, per-mode=symbol]{siunitx}
\usepackage[nomain,acronyms]{glossaries}
\usepackage{xspace}
\usepackage[capitalise, noabbrev]{cleveref}
\usepackage{caption}
\usepackage{subcaption}

\jvol{XX}
\jnum{XX}
\paper{8}
\jmonth{x/x}
\jname{IEEE Micro}
\pubyear{xxxx}

\graphicspath{{figures/}}

\newcommand{\riscv}{RISC-V\xspace}

\DeclareSIUnit\SFLOP{SPflop}
\DeclareSIUnit\DFLOP{DPflop}
\DeclareSIUnit\SFLOPs{\SFLOP\per\second}
\DeclareSIUnit\DFLOPs{\DFLOP\per\second}
\DeclareSIUnit\GSFLOPs{\giga\SFLOPs}
\DeclareSIUnit\GDFLOPs{\giga\DFLOPs}
\DeclareSIUnit\TSFLOPs{\tera\SFLOPs}
\DeclareSIUnit\TDFLOPs{\tera\DFLOPs}
\DeclareSIUnit\SFLOPsW{\SFLOP\per\second\per\watt}
\DeclareSIUnit\DFLOPsW{\DFLOP\per\second\per\watt}
\DeclareSIUnit\GSFLOPsW{\giga\SFLOPsW}
\DeclareSIUnit\GDFLOPsW{\giga\DFLOPsW}
\DeclareSIUnit\gateequivalent {GE}
\DeclareSIUnit\kGE {\kilo\gateequivalent{}}
\DeclareSIUnit\MGE {\mega\gateequivalent{}}

\newacronym{ast}{AST}{Abstract Syntax Tree}
\newacronym{os}{OS}{operating system}
\newacronym{hbm}{HBM}{High-Bandwidth Memory}
\newacronym{fpu}{FPU}{floating-point unit}
\newacronym{fp}{FP}{floating-point}
\newacronym{isa}{ISA}{Instruction Set Architecture}
\newacronym{ssr}{SSR}{Stream Semantic Register}
\newacronym{frep}{FREP}{Floating-point Repetition}
\newacronym{fma}{FMA}{Fused Multiply-Add}
\newacronym{dma}{DMA}{direct memory transfer}
\newacronym{sp}{SP}{single-precision}
\newacronym{dp}{DP}{double-precision}
\newacronym{fdsoi}{FD-SOI}{Fully-Depleted Silicon-on-Insulator}
\newacronym{numa}{NUMA}{non-uniform memory access}
\newacronym{simd}{SIMD}{single instruction, multiple data}
\newacronym{simt}{SIMT}{single instruction, multiple thread}
\newacronym{dvfs}{DVFS}{dynamic voltage and frequency scaling}
\newacronym{dnn}{DNN}{Deep Neural Network}
\newacronym{sve}{SVE}{scalable vector extension}
\newacronym{vliw}{VLIW}{very long instruction word}

\begin{document}

\sptitle{Department: Head}
\editor{Editor: Name, xxxx@email}

\title{Manticore: A 4096-core RISC-V Chiplet Architecture for Ultra-efficient Floating-point Computing}

\author{Florian Zaruba\textsuperscript{$*\dag$}, Fabian Schuiki\textsuperscript{$*\dag$}, Luca Benini\textsuperscript{$\dag\ddag$}}

\affil{
    {
        \textsuperscript{$\dag$}Integrated Systems Laboratory, ETH Zürich \quad
        \textsuperscript{$\ddag$}DEI, University of Bologna \quad
        \textsuperscript{$*$}equal contribution
    }
}
\markboth{Department Head}{Paper title}

\begin{abstract}
Data-parallel problems demand ever growing \gls{fp} operations per second under tight area- and energy-efficiency constraints. In this work we present \emph{Manticore}, a general-purpose, ultra-efficient chiplet-based architecture for data-parallel \gls{fp} workloads.
We have manufactured a prototype of the chiplet's computational core in Globalfoundries 22FDX process and demonstrate more than 5x improvement in energy efficiency on \gls{fp} intensive workloads compared to CPUs and GPUs. The compute capability at high energy and area efficiency is provided by Snitch clusters~\cite{zaruba2020snitch} containing eight small integer cores, each controlling a large \gls{fpu}.
The core supports two custom ISA extensions:
The \gls{ssr} extension elides explicit load and store instructions by encoding them as register reads and writes~\cite{schuiki2019stream}.
The \gls{frep} extension decouples the integer core from the \gls{fpu} allowing floating-point instructions to be issued independently.
These two extensions allow the single-issue core to minimize its instruction fetch bandwidth and saturate the instruction bandwidth of the \gls{fpu}, achieving \gls{fpu} utilization above 90\%, with more than 40\% of core area dedicated to the \gls{fpu}.
\end{abstract}

\maketitle

\glsresetall


\chapterinitial{Introduction}

Domains such as data-analytics, machine learning and scientific computing are dependent on increasing compute resources~\cite{openai2018aicompute}. Increasing technology node densities result in systems that are mainly limited by thermal design power and the most feasible way to increase the amount of active compute units is to design more energy-efficient architectures.
While many emerging architectures~\cite{jouppi2020domain}, especially in the machine learning domain, trade-off \gls{fp} precision for higher throughput and efficiency, algorithms such as stencils, linear differential equations require higher precision arithmetic. Domain-specific accelerators are a prominent example for how to leverage specialization~\cite{yang2019nervana}. Unfortunately, they are hard to adjust to algorithmic changes and tied to a specific application domain.

The trend in leading-edge general-purpose computer architectures paints a similar picture on the importance of increasing energy-efficiency. Two prominent examples of recent high-performance architectures are Fujitsu's A64FX~\cite{yoshida2018fujitsu} and NVIDIA's A100~\cite{nvidia2020amper}. Both systems strive to control their 32 (A64FX) and 16 (A100) wide multi-lane \gls{sp} data-path with as few instructions as possible. 


With the proposed Manticore system, we pursue a similar goal. We achieve this goal by pairing a simple, in-order, 32-bit \riscv integer core with a large \gls{fpu}. Two \gls{isa} extensions: \gls{ssr} and \gls{frep} make it possible for the single-issue integer core to saturate the bandwidth of its \gls{fpu}, achieving utilization higher than \SI{90}{\percent} for compute-bound kernels.


\section{Chiplet Architecture}
\label{sec:arch}

\begin{figure}
    \begin{center}
    \includegraphics[width=\linewidth]{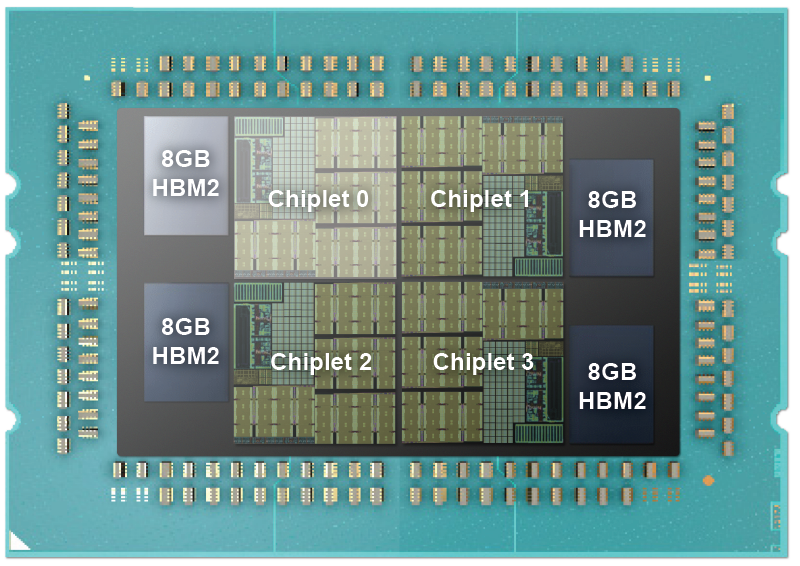}
    \end{center}
    \caption{
        Conceptual floorplan of the package.
        Shown are the arrangement of the chiplets and \glspl{hbm} on the interposer.
        Each chiplet has its own, private, \SI{8}{\giga\byte} \gls{hbm}.
        Chiplets interconnect via die-to-die serial links~\cite{vivet2020chiplet}.
    }
    \label{fig:floorplan_package}
\end{figure}

The proposed \emph{Manticore} architecture consists of four \SI{222}{\milli\meter\squared} (\SI{14.9 x 14.9}{\milli\meter}) 22\,nm chiplet dies on an interposer. Using chiplets improves yield and reduces cost. Each die has three short-range, multi-channel, in-package chip-to-chip links~\cite{vivet2020chiplet}, one to each sibling. They are used for inter-die synchronization and chiplet-to-chiplet \gls{numa}. Furthermore, each chiplet has access to a private \SI{8}{\giga\byte} \gls{hbm}. The conceptual floorplan is depicted in \cref{fig:floorplan_package}.

\begin{figure}
    \begin{center}
    \includegraphics[width=\linewidth]{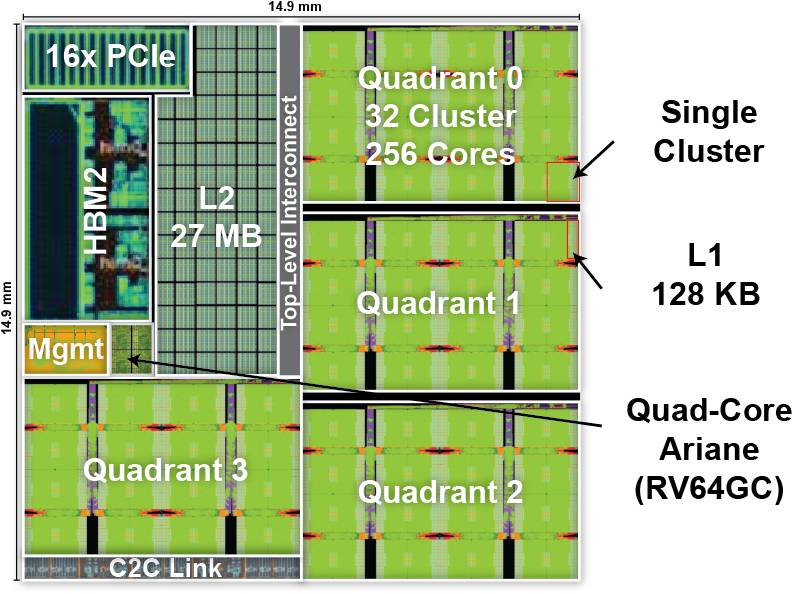}
    \end{center}
    \caption{
        Conceptual floorplan of an individual chiplet.
        Shown is the arrangement of individual cluster quadrants, interconnects, L2 memory, HBM2 controller, PCIe controller, and quad-core Ariane RV64GC system.
    }
    \label{fig:floorplan_chiplet}
\end{figure}

The chiplet (see~\cref{fig:floorplan_chiplet}) contains four quadrants, consisting of 32 clusters with eight cores each, which results in 1024 cores for all four quadrants on a chiplet. Furthermore, each chiplet contains four Ariane RV64GC management cores~\cite{zaruba2019cost} an \gls{hbm} (\SI{256}{\giga\byte\per\second}) controller, a \SI{27}{\mega\byte} of L2 memory, and a 16x PCIe endpoint (\SI{31.5}{\giga\byte\per\second}) for host communication as shown in \cref{fig:floorplan_chiplet}. 

The four Ariane management cores run a general-purpose operating system such as Linux and manage the Snitch clusters and program off-loading. The Manticore chiplet has enough silicon area to support \SI{27}{\mega\byte} on-chip shared L2 memory for critical data storage such as neural network weights or stencil kernels.

\subsection{Memory Hierarchy}
\label{sec:arch_mem}

\begin{figure}
    \begin{center}
    \includegraphics[width=\linewidth]{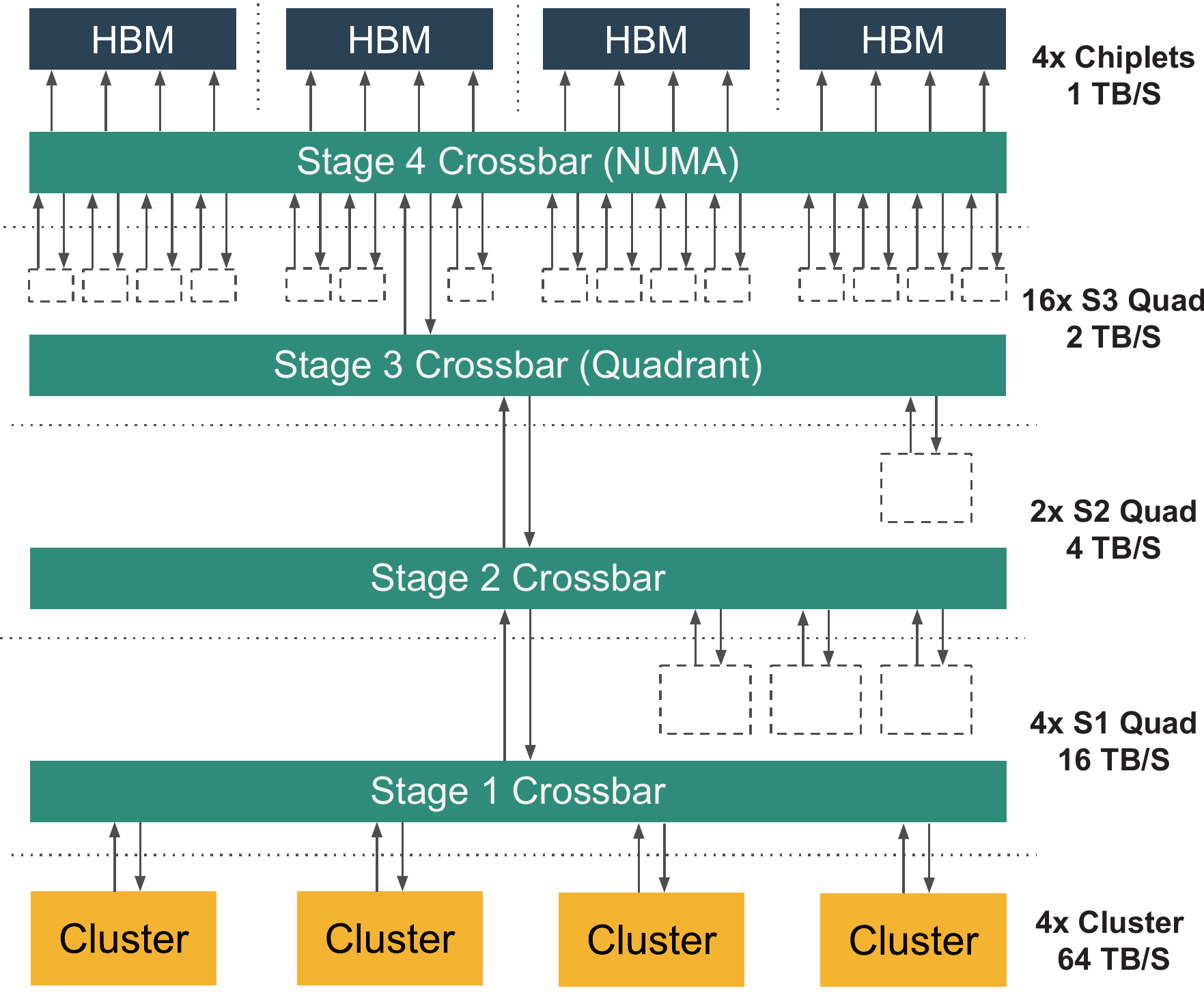}
    \end{center}
    \caption{
        Memory hierachy of the Manticore concept. Four cluster form a quadrant and share an uplink into the next stage. Four S1 quadrants form an S2 quadrant which share an uplink to the next stage. Two S2 quadrant form an S3 quadrant. Four S3 quadrants per chiplet share access to the \gls{hbm} memory.
    }
    \label{fig:memory_hierarchy}
\end{figure}

Each quadrant\footnote{The term quadrant is somewhat generic and does not necessarily imply four members (cores or lower stage quadrants), as the number of members can be adjusted to match the available bandwidth into the next stage, as for example, in stage three of our system} (see~\cref{fig:memory_hierarchy}) is further subdivided into multiple stages, in a tree-structure using an interconnect tuned for burst-based \gls{dma} accesses. Four clusters share an instruction cache and an uplink into the next stage. These four clusters have a high aggregate bandwidth of \SI{64}{\tera\byte\per\second} amongst each other and can perform low-latency, high-bandwidth intra-cluster data transfers. As shown in~\cref{fig:memory_hierarchy}, clusters share the uplink into the next higher stage, the bandwidth to the other S1 quadrants becomes smaller. Bandwidth is subsequently thinned as four S1 quadrants share an instruction cache and an uplink into the S2 quadrant and two S2 quadrants share an uplink into the S3 quadrant. In the last stage of hierarchy $16\times$ S3 quadrants, distributed over four chiplets (non-uniform memory access), share four \glspl{hbm} with an aggregated peak bandwidth of \SI{1}{\tera\byte\per\second}. This bandwidth thinning scheme allows us to have a very low diameter, low latency interconnect topology, which can sustainably saturate the \gls{hbm} bandwidth while being benign to floorplanning and physical design. The interconnect also allows for a very high cluster-to-cluster internal bandwidth, through multiple stages, which by far exceeds the bandwidth into the memory. With this model, we efficiently support cluster-to-cluster traffic, while, at the same time, fully loading the memory system.

\subsection{Compute Cluster}
\label{sec:arch_cluster}
    
\begin{figure}
    \begin{center}
    \includegraphics[width=\linewidth]{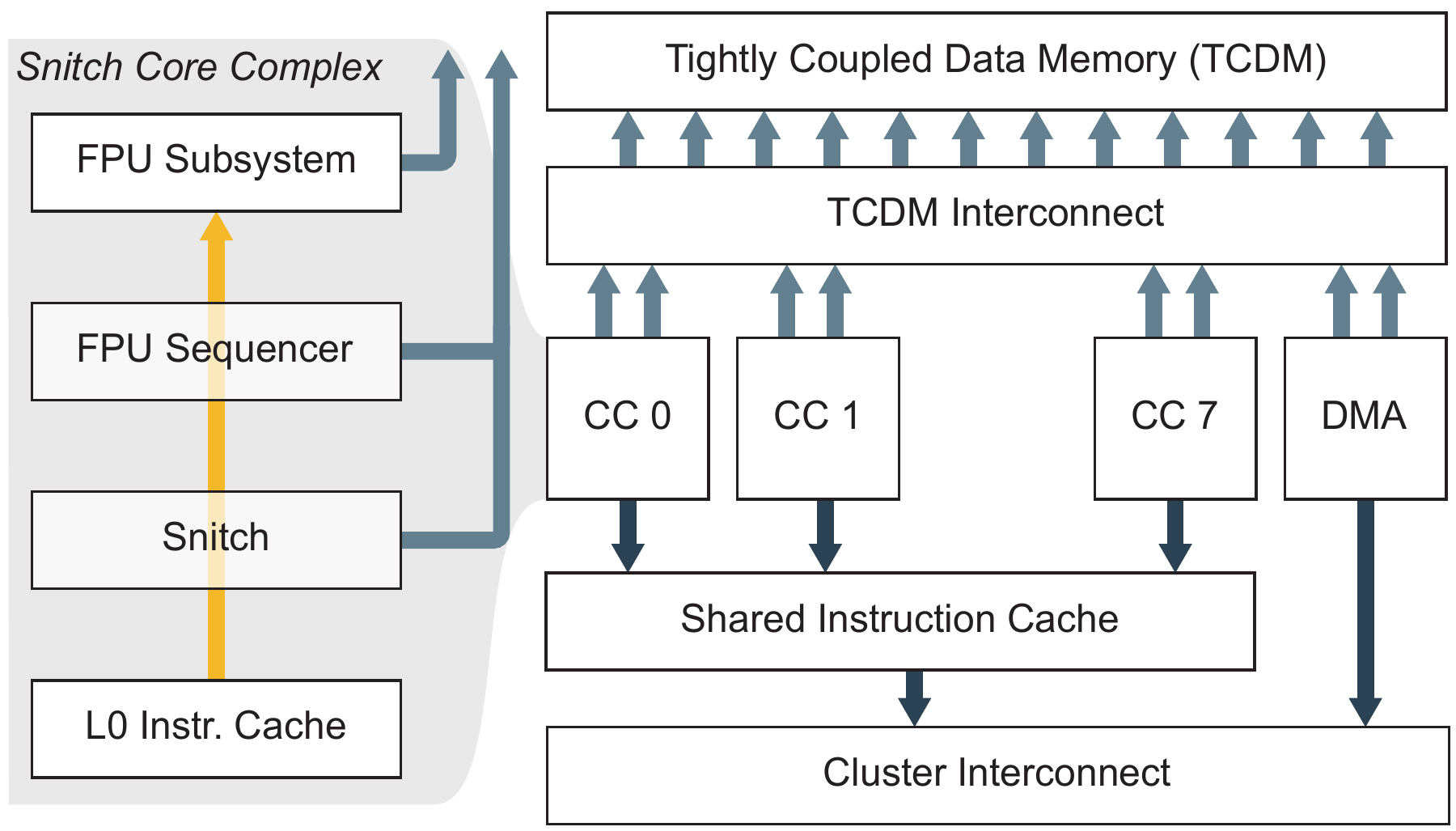}
    \caption{
        Simplified block diagram of a Snitch-based compute cluster. The core complex (CC) contains the integer core and the \gls{fpu} as well as necessary hardware for the \glspl{ssr} and \gls{frep}. The cluster contains eight core corplices which share an instruction cache and a tightly-coupled data memory. A \gls{dma} engines is used for efficient, bulk, data-movement.
    }
    \label{fig:snitch_cluster}
    \end{center}

\end{figure}

The compute cluster consists of eight small, \SI{22}{\kGE}, single-stage, 32 bit RISC-V processor cores~\cite{zaruba2020snitch} (see~\cref{fig:snitch_cluster}). Each Snitch core contains a \gls{dp} \gls{fpu}, which can be used to compute one \gls{dp} \gls{fma} operation or two \gls{sp} \glspl{fma} per cycle. When running at \SI{1}{\giga\hertz}, a cluster with eight Snitch cores is able to compute 16 \gls{dp} or 32 \gls{sp} flop, resulting in \SI{4}{\TDFLOPs} for the entire Manticore system. All eight cores have element-wise, low-latency, access into \SI{128}{\kilo\byte} tightly coupled and shared scratchpad memory. Moreover, a \gls{dma} engine is in charge of moving blocks of data into the scratchpad memory over a 512 bit, data bus. The cores are clocked at \SI{1}{\giga\hertz}, thus delivering more than \SI{4}{\tera\DFLOPs} peak compute per chiplet.


With this architecture, we achieve a very high compute/control ratio: \SI{44}{\percent} of the system consisting of compute units, another \SI{44}{\percent} spent on the L1 memory and just \SI{12}{\percent} of the area are spent on the control parts. 



\section{Programming}
\label{sec:prog}

We leverage two custom \riscv \gls{isa} extensions to achieve extremely high \gls{fp} utilization and efficiency: Xssr and Xfrep.

\subsection{Stream Semantic Registers (Xssr)}
\label{sec:prog_ssr}

\begin{figure}
    \begin{subfigure}[b]{0.48\textwidth}
    \centering
    \includegraphics[width=\linewidth]{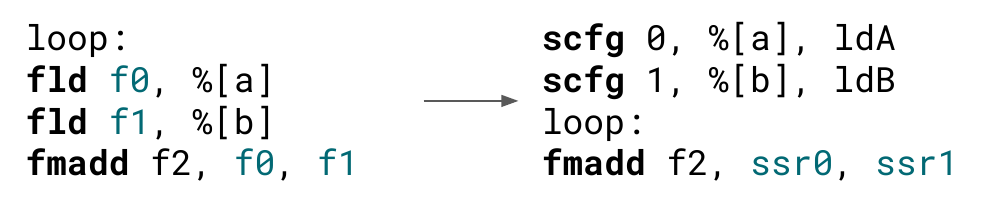}
    \caption{
        \emph{Left:} baseline simplified \riscv implementation, with address calculation and pointer increment omitted for brevity.
        \emph{Right:} \gls{ssr} implementation with memory loads encoded as reads from stream registers; additional stream configuration instructions required ahead of the loop.
    }
    \label{fig:snippet_ssr}
    \end{subfigure}
    \begin{subfigure}[b]{0.48\textwidth}
    \centering
    \includegraphics[width=\linewidth]{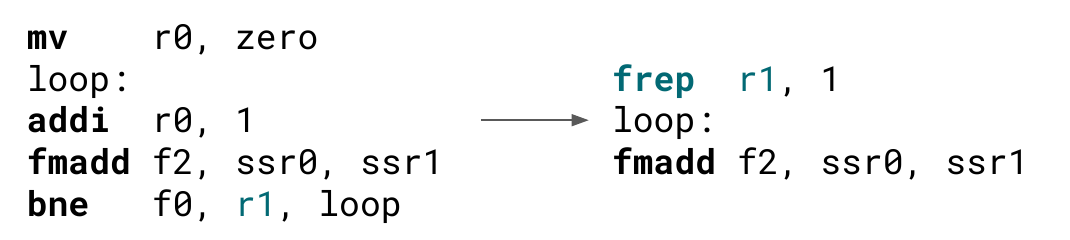}
    \caption{
        \emph{Left:} implementation with loop bookkeeping using baseline \riscv instructions.
        \emph{Right:} implementation with an \gls{frep} hardware loop, with all bookkeeping to occur implicitly in the hardware.
    }
    \label{fig:snippet_frep}
    \end{subfigure}
    \caption{
        The effect of \glspl{ssr} and \gls{frep} on the hot loop of a dot product kernel.
    }
\end{figure}

\Glspl{ssr} \cite{schuiki2019stream} offer a means to elide explicit load/store instructions in a program.
This is achieved by giving a subset of the processor core's registers \emph{stream semantics}.
When enabled, a read from such an \gls{ssr} is translated in hardware into a load from memory, and conversely, a register write becomes a store to memory.
Since an in-order single-issue core can only execute a single instruction every cycle, the presence of loads and stores in a hot loop of the program diminishes \gls{fpu} utilization significantly.
For example, consider a dot product, which has to issue two loads from memory for every \gls{fma} operation, as shown in \cref{fig:snippet_ssr}.
In this scenario, even if the loop is fully unrolled, we achieve at most 33\% \gls{fpu} utilization.
In theory, this allows the \gls{fpu} to be 100\% utilized, and even a simple processor can achieve >90\% utilization in many relevant kernels without resorting to complex and energy-inefficient wide issue superscalar or \gls{vliw} architectures~\cite{schuiki2019stream}. \glspl{ssr} offer a way to elide memory accesses and address computation in hot loops, which in many cases leaves no integer instructions in the loop body.

\subsection{Floating-Point Repetition (Xfrep)}
\label{sec:prog_frep}

The \gls{frep} \cite{zaruba2020snitch} extension implements a \gls{fpu}-only hardware loop.
Consider a dot product utilizing \glspl{ssr} for example, as shown in \cref{fig:snippet_frep}.
Besides the essential \gls{fma} operation running on the \gls{fpu}, the loop only consists of a trip count increment (\texttt{addi}) and a back-branch (\texttt{bne}).
This loop can be replaced by a \gls{frep} instruction, which loops a range of subsequent \gls{fp} instructions (one in this case) for a configurable number of times.
The \riscv \gls{isa} makes the integration of such an extension very straightforward as most instructions either operate entirely on integer or entirely on \gls{fp} registers. Only a handful, such as comparisons or moves between integer and \gls{fp} domains, exchange information from one domain to the other.
We leverage this separation and insert a micro-loop sequence buffer of 16 instructions between the Snitch core and the \gls{fpu}.
\Gls{frep} instructions configure this buffer to emit a range of buffered instructions multiple times into the \gls{fpu}, which essentially implements the hardware loop.
Since this happens entirely in the \gls{fpu} subsystem outside of the Snitch core, the core's integer pipeline can run in parallel, enabling a \emph{pseudo-dual-issue} mode of operation that would not be achievable with a traditional hardware loop.
This allows the core to perform non-trivial bookkeeping and address calculation while the \gls{fpu} is running, without incurring a reduction of the \gls{fpu} utilization.

\subsection{Typical SSR/FREP Execution}
\label{sec:prog_exec}

\begin{figure*}
    \centering
    \begin{subfigure}[t]{0.32\textwidth}
        \includegraphics[width=\textwidth, trim={8 8 8 8}, clip]{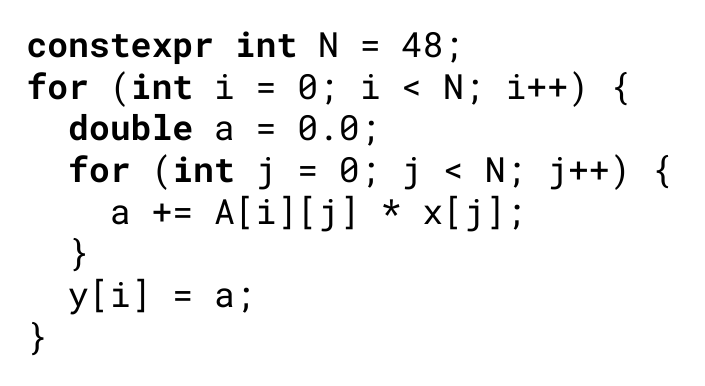}
        \caption{Reference C implementation with a square matrix \texttt{A} of fixed size 48.}
        \label{fig:snippet_matmul}
    \end{subfigure}
    \hfill
    \begin{subfigure}[t]{0.32\textwidth}
        \includegraphics[width=\textwidth]{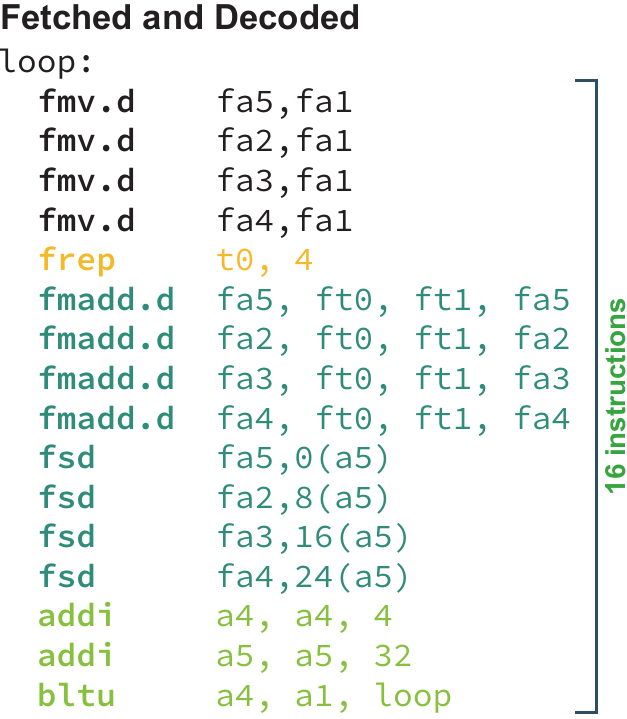}
        \caption{Resulting assembly implementation as stored in the binary and fetched/decoded by the processor core.}
        \label{fig:assembler_1}
    \end{subfigure}
    \hfill
    \begin{subfigure}[t]{0.32\textwidth}
        \includegraphics[width=\textwidth]{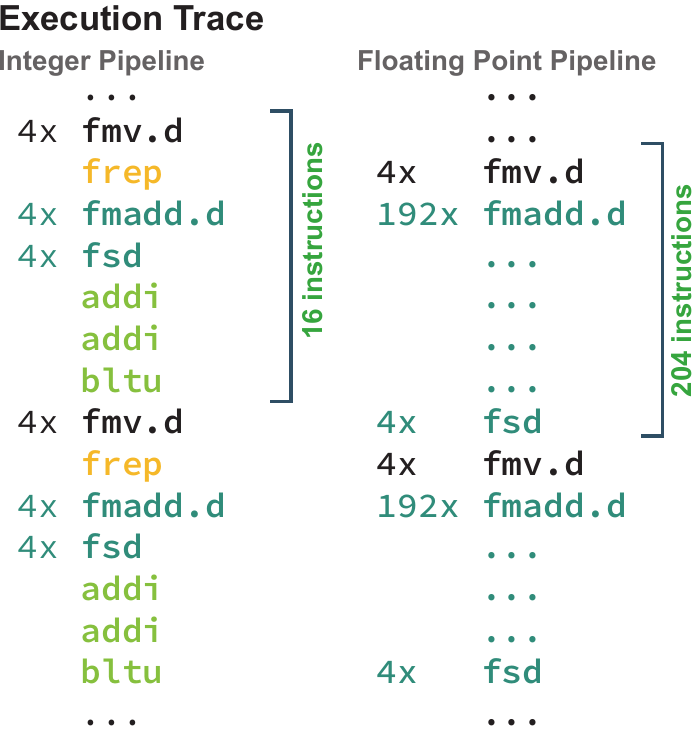}
        \caption{Execution traces of the integer pipeline (left) and the \gls{fp} pipeline (right).}
        \label{fig:assembler_2}
    \end{subfigure}
    \caption{
        Typical execution of a matrix-vector multiplication implementation leveraging the \gls{ssr} and \gls{frep} extensions.
        The 16 instructions are fetched and decoded once by the integer pipeline of the processor core (b), and expanded to 204 executed instructions in the \gls{fpu} (c).
    }
    \label{fig:plot_assembly}
\end{figure*}

As a concrete example, let us consider the matrix-vector multiplication operation shown in \cref{fig:snippet_matmul}.
A typical implementation leveraging Manticore's \gls{ssr} and \gls{frep} extensions is shown in \cref{fig:assembler_1}.
The address computation and memory accesses of \texttt{A} and \texttt{x} are entirely performed by the \glspl{ssr} \texttt{ft0} and \texttt{ft1}.
The inner loop is implemented using an \gls{frep} instruction and unrolled to compute four results in parallel in order to avoid pipeline stalls due to \gls{fpu} latency.
The outer loop is executed by the integer core. It stores the results (\texttt{fsd}), implements loop bookkeeping (\texttt{addi}, \texttt{bltu}), and initializes \texttt{a} (\texttt{fmv.d}).

As shown in \cref{fig:assembler_2}, the 16 instructions of the assembly implementation are fetched and decoded once by the integer pipeline of the processor core and expand to 204 executed instructions in the \gls{fpu} through the use of \gls{frep}.
This leaves 188 cycles for the integer pipeline for other tasks, such as preparing the next loop iteration or coordination of data movement.
In case no other work is required, the 16 instructions fetched over 204 cycles of execution amounts to roughly one instruction every 13 cycles, mitigating the von Neumann bottleneck by reducing instruction fetch bandwidth by more than one order of magnitude.
Since the \gls{fpu} can execute the loop iterations back-to-back and of the 204 instructions, 192 perform actual computation, this kernel can achieve up to 94\% \gls{fpu} utilization.


\section{Prototype}
\begin{figure*}
    \begin{center}
    \includegraphics[width=\linewidth]{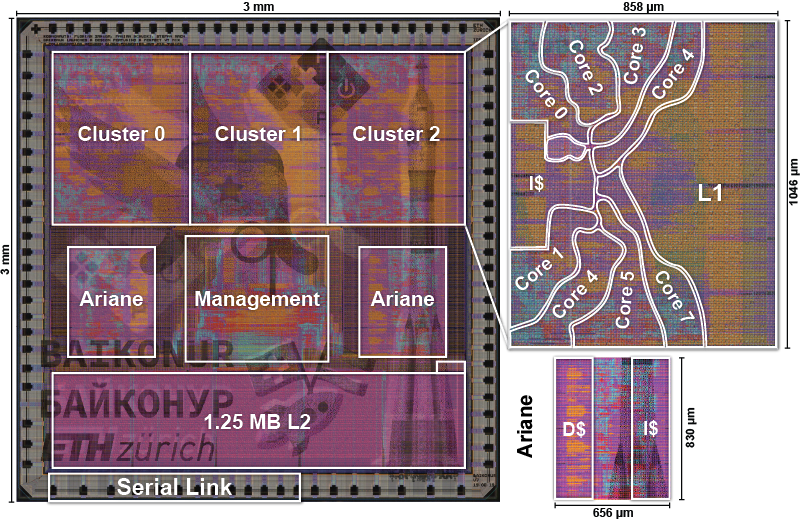}
    \end{center}
    \caption{
        Floorplan of the prototype silicon. The two Ariane cores as well as the Snitch cluster have been designed hierarchically. The core's follow a star-shaped layout around the shared instruction cache.
    }
    \label{fig:floorplan_prototype}
\end{figure*}


A \SI{3 x 3}{\milli\meter} prototype containing the logic core of the chiplet architecture was manufactured and characterized using the Globalfoundries 22FDX technology.
The prototype in \cref{fig:floorplan_prototype} contains three Snitch clusters with eight cores (each configured with \SI{8}{\kilo\byte} L1 instruction cache and \SI{128}{\kilo\byte} L1 data memory organized in 32 banks), a dual-core Ariane (with \SI{16}{\kilo\byte} L1 instruction cache and \SI{32}{\kilo\byte} data cache), \SI{1.25}{\mega\byte} L2 memory, and a \SI{400}{\mega\hertz}, double data-rate, \SI{2.56}{\giga\bit\per\second}, digital-only chip-to-chip link. 



\section{Silicon Performance}
\label{sec:perf}

\subsection{Efficiency}
\label{sec:perf_eff}

\begin{figure}
    \centering
    \includegraphics[width=\linewidth]{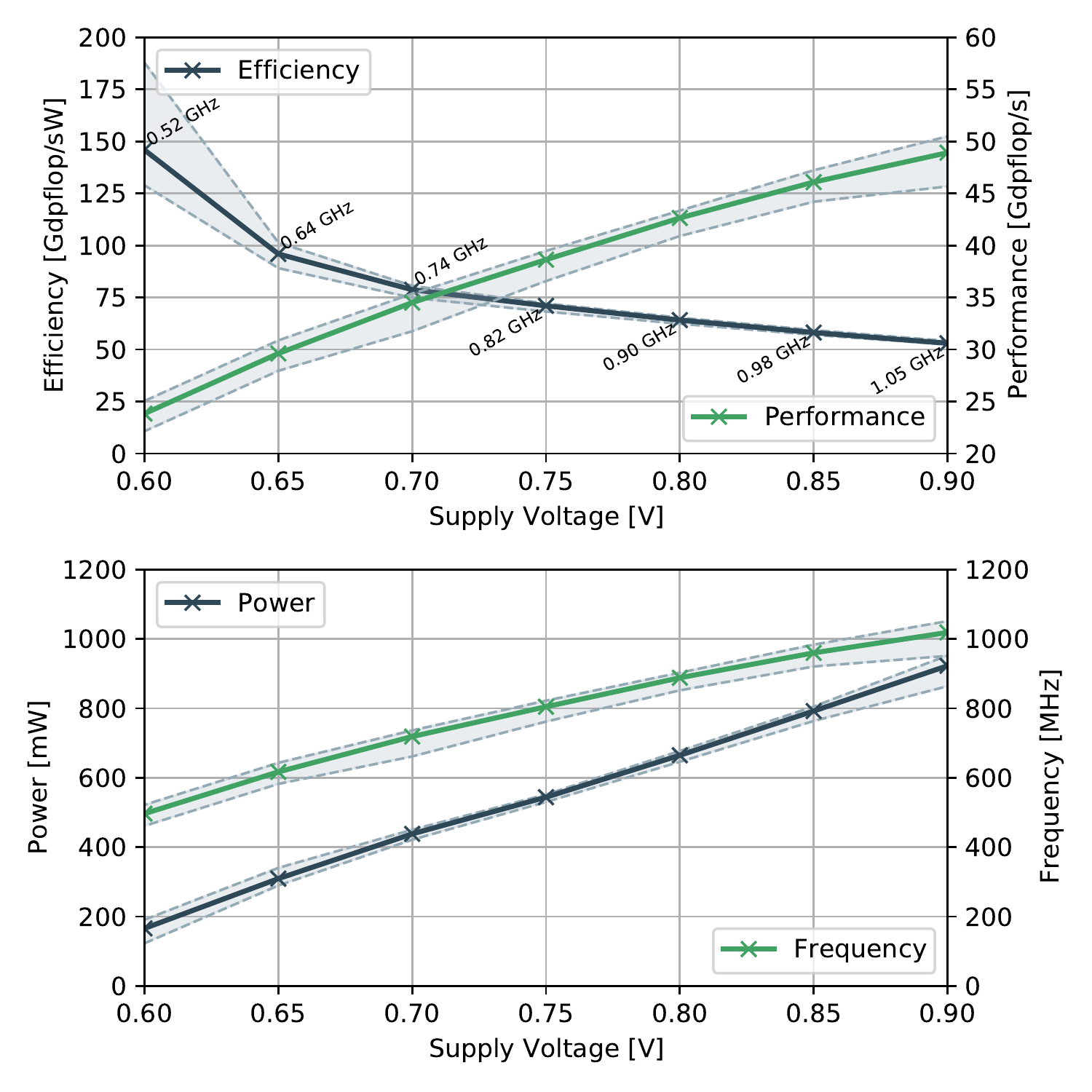}
    \caption{
        Compute performance, energy efficiency, speed, and power consumption for different operating voltages.
        Measured on the prototype silicon across eight sample dies.
        Cores performing matrix multiplications, at 90\% FPU utilization.
        Performance and efficiency doubles across range.
    }
    \label{fig:plot_perf_eff}
\end{figure}

We measured the speed and power consumption of the prototype silicon under representative workloads and operating conditions.
\cref{fig:plot_perf_eff} shows the \gls{dp} performance and energy efficiency achieved when executing parallel workloads on the 24 cores of our prototype 22\,nm silicon, for different operating voltages.
The chips offer a wide range of operating points and choices for performance/efficiency trade-offs, which we leverage through \gls{dvfs} based on the current workload's operational intensity.
This allows us to essentially adjust the roofline of the system to match the current workload.
In \emph{high-performance mode} running over \SI{1}{\GHz} at \SI{0.9}{\volt} VDD, our architecture achieves a peak performance of \SI{54}{\giga\DFLOPs} across 24 cores and a compute density of up to \SI{20}{\giga\DFLOPs\per\milli\meter\squared}, which translates to \SI{9.2}{\tera\DFLOPs} across a full 4096 cores.
In \emph{max-efficiency mode} running at \SI{0.5}{\GHz} at \SI{0.6}{\volt} VDD, our architecture achieves an industry-leading efficiency of \SI{188}{\giga\DFLOPsW}, while still delivering a respectable \SI{25}{\giga\DFLOPs} across 24 cores, which translates to \SI{4.3}{\tera\DFLOPs} across a full 4096 cores.

\subsection{Roofline}
\label{sec:perf_roofline}

\begin{figure}
    \centering
    \includegraphics[width=\linewidth]{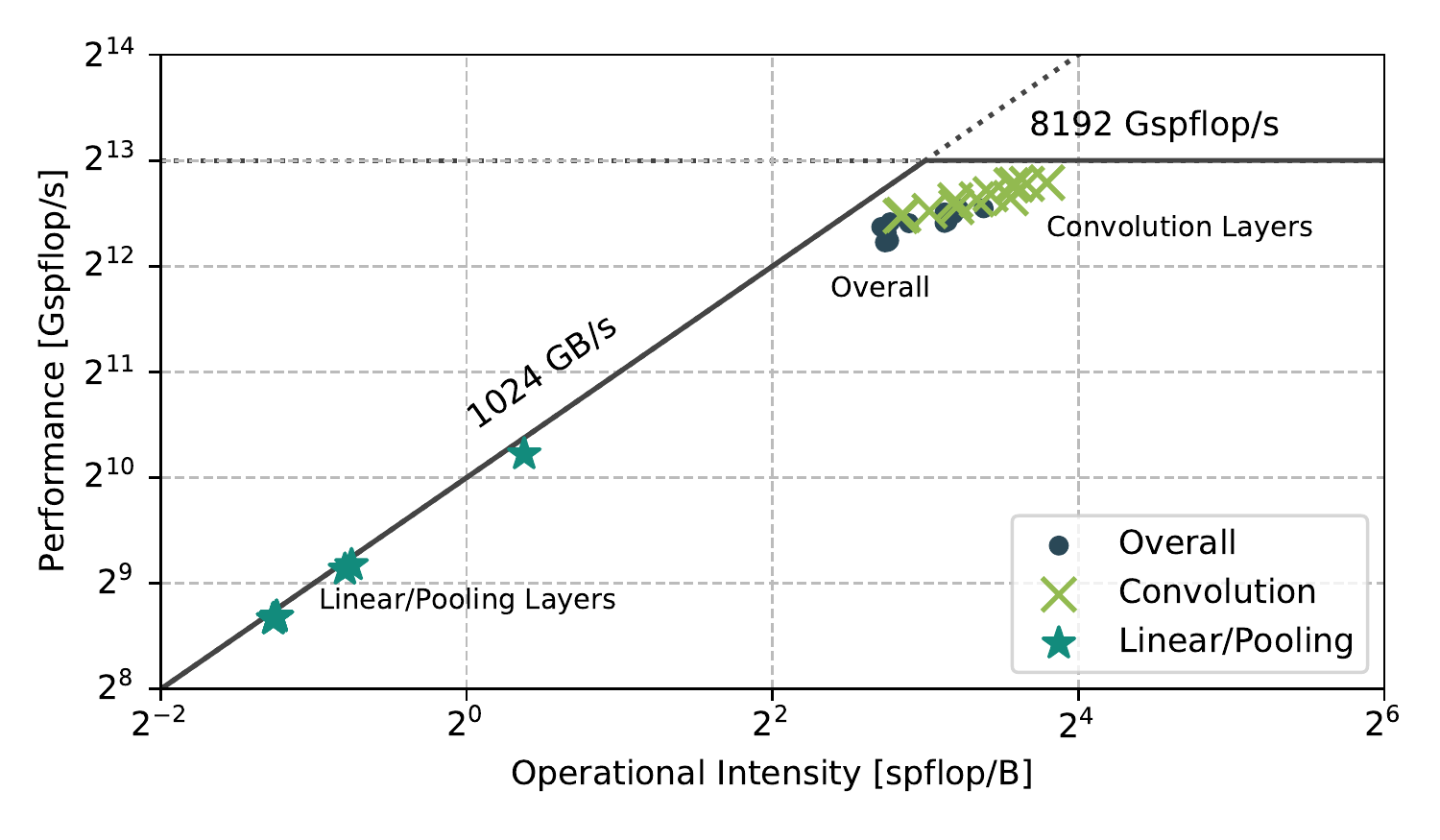}
    \caption{
        Performance roofline plot of DNN training workloads.
        We group convolutions and linear/pooling layers to indicate performance in the compute- and memory-bound regions, respectively.
        The Manticore architecture is very efficient at tracking the performance and bandwidth roofline, with a detachment down to 5\% for low-intensity and 14\% for high-intensity optimized kernels.
    }
    \label{fig:plot_dnn_roofline}
\end{figure}

To assess the performance of the manufactured silicon, we analyzed workloads from training steps of a set of \glspl{dnn}.
\cref{fig:plot_dnn_roofline} shows the roofline plot of our architecture across a full training step.
We estimate full-system performance based on cycle-accurate simulation of a smaller instantiation of the hardware, combined with an architectural model of the full system and measured performance characteristics of the prototype silicon.
The compute-bound convolution layers in the workload reach >80\% of the system's peak performance, and the proximity to the point of inflection of the roofline indicates a balanced utilization of the hardware capabilities.
The memory-bound linear and pooling layers reach >90\% of the system's peak bandwidth.
Since \gls{dnn} workloads tend to be dominated by the convolution layers, the overall performance, which considers all layers, is almost identical to the convolution performance.
Overall we observe that the Manticore architecture is very efficient at tracking the performance and bandwidth roofline, with a detachment down to \SI{5}{\percent} for low-intensity and \SI{14}{\percent} for high-intensity optimized kernels.
The worst-case detachment of \SI{34}{\percent} from the roofline is encountered in the intermediate region around the point of inflection, where intuitively, the aggregate bandwidth pressure on the L1 TCDM is highest due to the DMA and the compute units both operating at capacity and banking conflicts more frequently stall L1 memory accesses.

\subsection{Applications}
\label{sec:perf_apps}

\begin{figure}
    \centering
    \includegraphics[width=\linewidth]{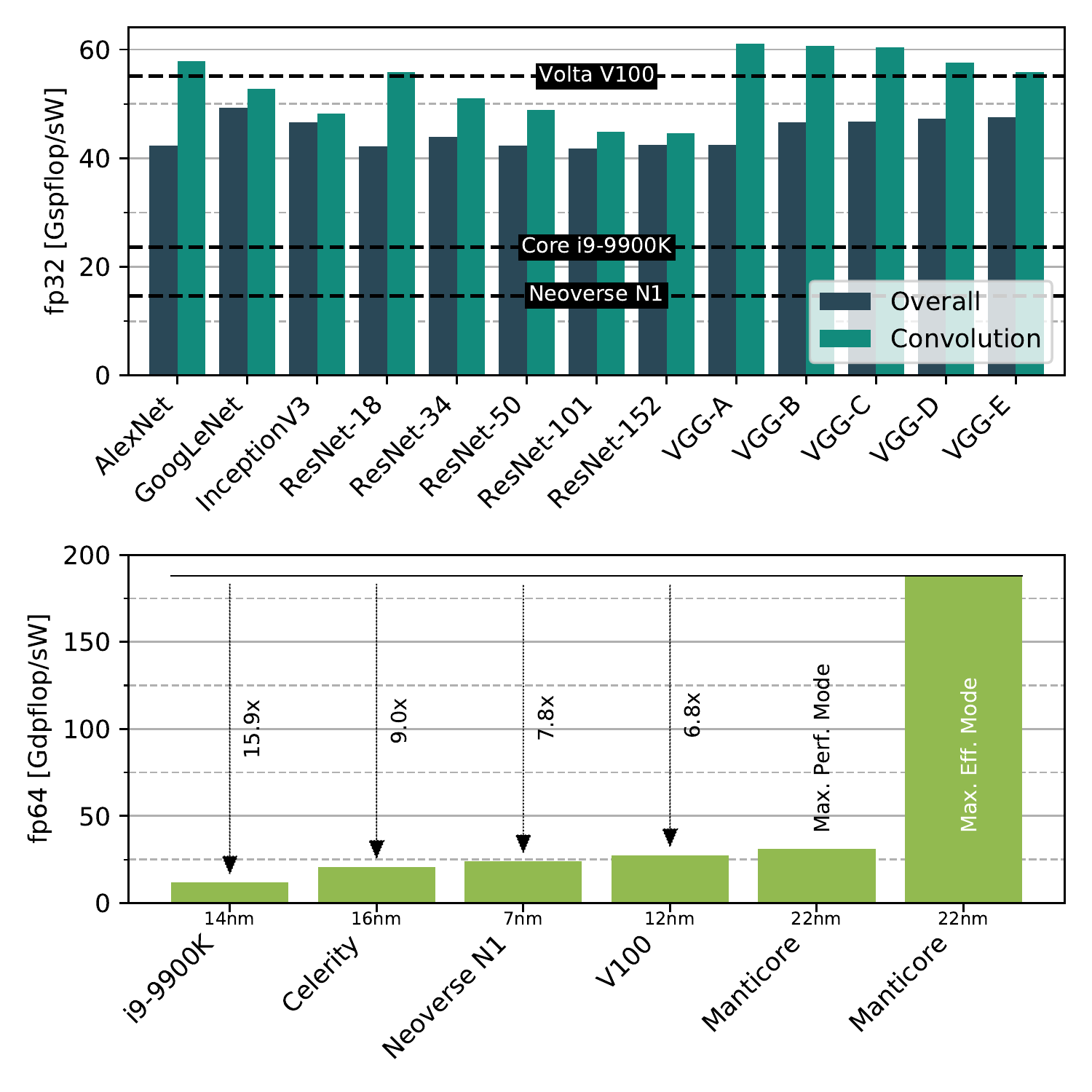}
    \caption{
        \emph{Top:}
        \Gls{sp} energy efficiency on full DNN training step, overall and specifically on the convolution layers.
        \emph{Bottom:}
        \Gls{dp} energy efficiency on linear algebra (assuming 90\% of peak performance);
        Manticore shown for maximum performance and maximum efficiency operating points.
    }
    \label{fig:plot_dnn_eff}
\end{figure}

\cref{fig:plot_dnn_eff} shows the \gls{sp} energy efficiency achieved in a DNN training step overall, and on the compute-bound convolutions specifically, across a variety of networks, and the industry-leading \gls{dp} efficiency on linear algebra.
On \gls{sp} \gls{dnn} training workloads, Manticore's actual efficiency is competitive with the V100 GPU's peak efficiency and outperforms the Core i9-9900K CPU by 2$\times$ and the Neoverse N1~\cite{christy2020neoverse} by 3$\times$.
On \gls{dp} workloads, Manticore outperforms a V100 GPU's peak efficiency by 6$\times$, the N1 by 7$\times$, the Celerity \riscv CPU by 9$\times$, and the Core i9-9900K CPU by 15$\times$.
Our architecture achieves this despite these chips having a substantial technology advantage due to their 7\,nm, 12\,nm, and 14\,nm FinFET processes.
Regarding the A100 GPU, our initial estimates based on data published by Nvidia~\cite{nvidia2020amper} suggest that it achieves a 25\% improvement on \gls{sp} and \gls{dp} over the V100 in terms of speed at similar power consumption.
This indicates that Manticore has just 25\% lower efficiency on \gls{sp} than A100, but outperforms it on \gls{dp} by 5$\times$, despite the A100's significant 7\,nm FinFET technology advantage.
Manticore delivers significantly higher peak \gls{fp} performance than comparable \riscv architectures~\cite{ajayi2017celerity} in 16\,nm.





\subsection{Acknowledgment}

This project was supported in part by the European Union's H2020 program grant agreement numbers 826647 (European Processor Initiative-EPI) and 732631 (Open Transprecision Computing - "OPRECOMP").

\bibliographystyle{IEEEtran}
\bibliography{IEEEabrv,bibliography}

\begin{IEEEbiography}{Florian Zaruba}{\,}received his BSc degree from TU Wien in 2014 and his MSc from the Swiss Federal Institute of Technology Zurich in 2017. He is currently pursuing a PhD degree with the Digital Circuits and Systems group of Luca Benini. Contact him at zarubaf@iis.ee.ethz.ch.
\end{IEEEbiography}

\begin{IEEEbiography}{Fabian Schuiki}{\,}received his BSc and MSc degree in electrical engineering from ETH Zürich, in 2014 and 2016, respectively.  He is currently pursuing a PhD degree with the Digital Circuits and Systems group of Luca Benini. Contact him at fschuiki@iis.ee.ethz.ch.
\end{IEEEbiography}

\begin{IEEEbiography}{Luca Benini}{\,}holds the Chair of Digital Circuits and Systems at ETHZ and is Full Professor at the Universita di Bologna. He is a Fellow of the ACM and a member of the Academia Europaea. Contact him at lbenini@iis.ee.ethz.ch.
\end{IEEEbiography}

\end{document}